\documentclass[a4paper,11pt]{article}
\pdfoutput=1 % if your are submitting a pdflatex (i.e. if you have
\usepackage{jcappub} % for details on the use of the package, please
\usepackage[T1]{fontenc} % if needed

\title{\boldmath Deblurring the early Universe: reconstruction of primordial power spectrum from \textit{Planck} CMB using image analysis techniques}

\author[a]{Wuhyun Sohn,}
\author[a,b,1]{Arman Shafieloo,\note{Corresponding author.}}
\author[c,d,e]{Dhiraj Kumar Hazra}

\affiliation[a]{Korea Astronomy and Space Science Institute, Daejeon 34055, South Korea}
\affiliation[b]{University of Science and Technology, Daejeon 34113, South Korea}
\affiliation[c]{The Institute of Mathematical Sciences, HBNI, Chennai  600113, India}
\affiliation[d]{Homi Bhabha National Institute, Training School Complex, Anushakti Nagar, Mumbai 400094, India}
\affiliation[e]{INAF/OAS Bologna, Osservatorio di Astrofisica e Scienza dello Spazio, Area della ricerca CNR-INAF, via Gobetti 101, I-40129 Bologna, Italy}

\emailAdd{wuhyun@kasi.re.kr}
\emailAdd{shafieloo@kasi.re.kr}
\emailAdd{dhiraj@imsc.res.in}

\abstract{While the simplest inflationary models predict the primordial perturbations to be near scale-invariant, the primordial power spectrum (PPS) can exhibit oscillatory features in many physically well-motivated models. We search for hints of such features via free-form reconstructions of the PPS based on \textit{Planck} 2018 CMB temperature and polarization anisotropies. In order to robustly invert the oscillatory integrals and handle noisy unbinned data, we draw inspiration from image analysis techniques. In previous works, the Richardson-Lucy deconvolution algorithm for deblurring images has been modified for reconstructing PPS from the CMB temperature angular power spectrum. We extensively develop the methodology by including CMB polarization and introducing two new regularization techniques, also inspired by image analysis and adapted for our cosmological context. Regularization is essential for improving the fit to the temperature and polarization channels (TT, TE and EE) simultaneously without sacrificing one for another. The reconstructions we obtain are consistent with previous findings from temperature-only analyses. We evaluate the statistical significance of the oscillatory features in our reconstructions using mock data and find the observations to be consistent with having a featureless PPS. The machinery developed here will be a complimentary tool in the search for features with upcoming CMB surveys. Our methodology also shows competitive performance in image deconvolution tasks, which have various applications from microscopy to medical imaging.}

\begin{document}
\maketitle
\flushbottom

\section{Introduction}
\label{sec:introduction}

The structure of the Universe today originates from the perturbations present in the early Universe, whose statistical properties are well summarised by their power spectrum. The simplest models of inflation predict the primordial scalar power spectrum to be near-scale-invariant only with a small red tilt, characterised by its amplitude, $A_\mathrm{s}$, and tilt, $n_\mathrm{s}$ (see e.g. \cite{Martin2014EncyclopdiaInflationaris,Meerburg2019PrimordialNon-Gaussianity,Achucarro2022Inflation:Observations} for reviews). CMB observations to date have placed precise estimates on these two parameters \cite{PlanckCollaboration2018parameters} and are consistent with a near scale-invariant primordial power spectrum (PPS) \cite{PlanckCollaboration2018inflation}.

However, there are numerous physically well-motivated models of inflation that predict deviations away from a near-scale-invariant spectrum. Among them are models predicting oscillations in the PPS caused by, e.g. sharp features in the inflationary potential \cite{Starobinsky1992SpectrumPotential,Adams2001InflationaryStep,Covi2006InflationFuture} and resonance \cite{Chen2008GenerationInflation,Flauger2010OscillationsInflation,Flauger2011ResonantNon-Gaussianity,Aich2013OscillationsDatasets} (see \cite{Chen2010,Slosar2019ScratchesFluctuations,Achucarro2022Inflation:Observations} for reviews). A detection or non-detection of such features in the power spectrum would therefore constitute an illuminating probe of inflationary physics.
Furthermore, detecting such features may have implications for cosmological parameter estimation and the tension between the early and late-time probes in cosmology \cite{Hazra2022OneCosmology,DiValentino2021InSolutions,Abdalla2022CosmologyAnomalies}.

There have been two broad classes of approaches in the literature for studying deviations from the power-law spectrum. The first considers minimally parametric extensions to the power-law spectrum. A number of different approaches have been considered, including studying the scale variation of the tilt by introducing a running and/or running of the running \cite{PlanckCollaboration2013inflation,PlanckCollaboration2015inflation,PlanckCollaboration2018inflation}, oscillatory template-based analyses \cite{Ichiki2010CosmicSpectrum,Meerburg2012WMAP7Spectrum,Meerburg2014SearchingApproach,Meerburg2014SearchingData,Meerburg2016JointEstimation,Fergusson2015CombiningFeatures,Fergusson2015PolyspectraData,Hu2015SearchingSpectra}, the parametrization of PPS amplitudes at some nodes (or `knots') in Bayesian analysis with linear interpolations \cite{Bridle2003ReconstructingSpectrum,Bridges2009BayesianSpectrum} or cubic smoothing splines \cite{Sealfon2005spline,Verde2008spline,Peiris2010spline} in between the nodes whose locations may also vary \cite{Vazquez2012ModelSpectrum,Aslanyan2014TheSpectrum,Abazajian2014TheSpectrum,PlanckCollaboration2015inflation,PlanckCollaboration2018inflation,Handley2019BayesianData}, and the inclusion of PCA modes of the Fisher matrix \cite{Leach2006PCA}.

The second group of methods takes an opposite approach; they reconstruct the functional form of the PPS, either directly or by introducing a large number of degrees of freedom in the process. A suitable regularization process is often required to control overfitting for which a number of approaches have been proposed. Some early works made use of wavenumber bins \cite{Wang1999CosmologyModels,Hannestad2001ReconstructingData,Hannestad2004ReconstructingAlgorithm,Elgaroy2002}. The cosmic inversion method \cite{Matsumiya2002CosmicAnisotropy,Matsumiya2003CosmicData,Kogo2004ReconstructingMethod} solves for the PPS directly using differential equations derived from the cosmological perturbation theory. Free-form reconstructions can be obtained by maximising the likelihood subject to some roughness penalties on the function values \cite{Tegmark2002SeparatingBox}, their derivatives \cite{Tocchini-Valentini2005ASpectrum,Tocchini-Valentini2006Non-parametricData,Nagata2009Band-powerMethod,Hunt2014ReconstructionSets,Hunt2015SearchDatasets} (Tikhonov regularization), or their second derivatives \cite{Gauthier2012ReconstructingCMB,PlanckCollaboration2013inflation,PlanckCollaboration2015inflation,PlanckCollaboration2018inflation} (analogous to smoothing splines). Localised features have been studied using wavelet basis functions \cite{Mukherjee2003DirectSpectrum,Mukherjee2005PrimordialReconstruction}. A direct inversion using Singular Value Decomposition on the transfer kernel \cite{Nicholson2010ReconstructionInversion} and Modified Richardson-Lucy algorithm (MRL) \cite{Shafieloo2004MRL,Shafieloo2007FeaturesAnalysis,Shafieloo2008EstimationData,Nicholson2009ReconstructionExperiments,Gibelyou2010DetectabilityDistribution,Hamann2010FeaturesAnalysis,Hazra2013wmapMRL,Hazra2013CosmologicalSpectrum,Hazra2014PlanckMRL,Chandra2021PrimordialSpectrum} can also be grouped into this category.

The original authors of MRL drew inspiration from an analogous task in image analysis: image deblurring, or deconvolution. An observed image can be described by the convolution of the original signal with an imaging system's point spread function (PSF), or the imaging system's response to a point source. The Richardson-Lucy algorithm \cite{Richardson1972Bayesian-BasedRestoration,Lucy1974AnDistributions} is an iterative algorithm which enables reconstruction of the underlying image given the observed image and the PSF. This technique has been used extensively in astronomical imaging and was first brought into the cosmological context in \cite{Baugh1993TheFunction,Baugh1994TheSpectrum} and to CMB analysis in  \cite{Shafieloo2004MRL}. Analogously to image analysis, where the instrumental PSF acts to blur the image of an object, in cosmology, the CMB transfer functions act to evolve and project the primordial perturbations. The adapted algorithm, dubbed MRL, solves the inverse problem of reconstructing the primordial power spectrum $P_\zeta (k)$ from the observed CMB angular power spectrum $C_\ell$ given the transfer functions. After a suitable smoothing process which reduces the noise fitting, we are able to obtain a form of PPS which may contain hints of features \cite{Hamann2010FeaturesAnalysis,Hazra2013wmapMRL,Hazra2014PlanckMRL,Chandra2021PrimordialSpectrum}.

In this work, we build on previous MRL analyses, making two major modifications. First, we introduce two different regularization methods to reduce noise amplification. The first of the two, dubbed regularized MRL (RegMRL), combines the diffusion equation with MRL to let the features in the reconstructions diffuse away unless they are strongly favoured by the data. The second method adopts the total variation (TV) regularization which acts to minimize the number of non-zero variations. Moreover, we test both of these methods on real-world images, comparing with some other existing regularized Richardson-Lucy deconvolution techniques in image analysis: the Tikhonov-Miller \cite{Katsaggelos1991AAlgorithm,Conchello1996FastMicroscopy} and TV regularization \cite{Dey2004ARegularization}.

Second, we extend the formalism to include \textit{Planck} CMB polarization data, improving on our previous temperature-only analyses utilising WMAP \cite{Hazra2013wmapMRL} and \textit{Planck} \cite{Hazra2014PlanckMRL} data. In doing so, we seek free-form PPS which improve the goodness of fit to each of the CMB TT, TE and EE spectra simultaneously, not overfitting the TT at the cost of others. The regularization methods introduced in this work enable this by controlling the noise amplification from MRL iterations.

\section{Methods}
\label{sec:methodology}

\subsection{Modified Richardson-Lucy deconvolution algorithm}

The CMB angular power spectrum can be computed from the primordial curvature power spectrum $P_\zeta (k)$ by
\begin{align}
	C_\ell^{XY} = 4\pi \int d(\ln k) ~ \Delta^X_\ell(k) \Delta^Y_\ell(k) P_\zeta (k),
\end{align}
where $\Delta^X_\ell(k)$ is the CMB transfer function obtained from Boltzmann solvers such as CAMB \cite{Lewis2000camb} and CLASS \cite{Blas2011class}. Here $X=$ T or E, which correspond to the CMB temperature and E-mode polarization, respectively.  The dimensionless power spectrum is parameterised assuming a power-law form in the \textit{Planck} analysis so that $P_\zeta (k) = A_\mathrm{s} (k/k_\mathrm{pivot})^{n_\mathrm{s} - 1}$, where $A_\mathrm{s}$ and $n_\mathrm{s}$ indicate the scalar amplitude and tilt, respectively \cite{PlanckCollaboration2018parameters,PlanckCollaboration2018power,PlanckCollaboration2018inflation}. The pivot scale is set to $k_\mathrm{pivot}=0.05 \mathrm{Mpc}^{-1}$ following the convention in \cite{PlanckCollaboration2018parameters}.

We discretise the $k$ space using a one-dimensional grid (or bins) provided by CAMB so that
\begin{align}
	C_\ell^{XY} = \sum_k G_{\ell k}^{XY} P_k.	\label{eqn:Cl_equals_Glk_Pk}
\end{align}
The matrix $G^{XY}_{\ell k}$ depends on the cosmological parameters and captures relevant information about the evolution and projection of the CMB anisotropies. Note that the integration weights for the $k$ grid (bins) are also included in $G_{\ell k}$.

The relation \eqref{eqn:Cl_equals_Glk_Pk} can be used to reconstruct $P_\zeta (k)$ from the observed $C_\ell$. However, due to the oscillatory nature of the integrals appearing in \eqref{eqn:Cl_equals_Glk_Pk} and hence in $G_{\ell k}$, a direct inversion is non-trivial. MRL algorithm \cite{Shafieloo2004MRL} reconstructs $P_\zeta (k)$ from the data $C^\mathrm{obs}_\ell$ iteratively through 
\begin{align}
	P^{(i+1)}_k = P^{(i)}_k + P^{(i)}_k \sum_\ell \frac{C^\mathrm{obs}_\ell - C^{(i)}_\ell}{C^{(i)}_\ell} \tilde{G}_{\ell k}  \label{eqn:MRL_iteration} 
	\tanh^2 \left[ Q^{(i)}_\ell \left( C^\mathrm{obs}_\ell - C^{(i)}_\ell \right)  \right],   
\end{align}
where
\begin{align}
    C^{(i)}_\ell &\equiv \sum_k G_{\ell k} P^{(i)}, \\
    Q^{(i)}_\ell &\equiv \sum_{\ell'} \mathrm{Cov}^{-1}(\ell, \ell') \left( C^\mathrm{obs}_{\ell'} - C^{(i)}_{\ell'} \right), \\
    \tilde{G}_{\ell k} &\equiv G_{\ell k}\ /\ \sum_{\ell'} G_{\ell' k}.
\end{align}
At the $i$th iteration, $P_\zeta^{(i)}(k)$ is updated so that $C_{\ell}^{(i)}$ approaches closer to the observed data $C^\mathrm{obs}_{\ell}$. The $\tanh$ term appearing in \eqref{eqn:MRL_iteration} provides weights according to the noise covariance matrix; the points with more significant deviations away from the data are weighted more than others. By using a full covariance matrix we also include non-diagonal correlation information.

MRL provides iteratively solves the inverse problem \eqref{eqn:Cl_equals_Glk_Pk} in a non-linear manner, not necessarily minimising a particular loss function but gradually improving the fit to the data as iterations pass. We stop at the iteration when a desired condition, such as convergence, is satisfied. MRL reconstructions tend to converge within $O(100)$ iterations \cite{Shafieloo2004MRL}.

Due to the flexibility inherent in a free-form reconstruction, MRL can find a series of plausible solutions to any data by overfitting the noise. The reconstruction result should be interpreted with care. First, regularization is necessary to mitigate noise fitting and will be discussed in the following sections. Second, we note that we are \textit{not} proposing the reconstructed spectrum as a new `model' better than the power-law PPS of $\Lambda$CDM, since having more degrees of freedom can always improve the fit to data. Instead, we use the free-form reconstruction as a `catch-all' method to look for features that may be hiding in the data, aided with some statistical analysis using simulated data.

\subsection{Combining polarization and temperature data}

In this work, we use the unbinned, coadded $C_{\ell}$s and their covariance matrix from version 12.5 of the cleaned CamSpec likelihood \cite{Efstathiou2021AMaps}, which is a re-analysis of the \textit{Planck} 2018 data \cite{PlanckCollaboration2018power}. The range in $\ell$ for the TT, TE and EE angular power spectrum data used are $[2,2500]$, $[30,2500]$ and $[30,2000]$, respectively. The cleaned CamSpec 12.5 likelihood utilises 80\% of the sky in the temperature and polarization maps, which is the largest fraction compared to other likelihoods. This likelihood does not include the 100~GHz auto-correlation spectrum in the analysis and uses the 545~GHz map to clean the 143~GHz and 217~GHz maps. In order to obtain the foreground ${\cal C}_\ell$'s we find the best fit to the Planck's TTTEEE + low-$\ell$ + Simall data using BOBYQA \cite{Powell2009TheDerivatives.} minimization. The best-fit foregrounds were then subtracted from the spectra. From the maximum likelihood estimate, the co-added TT spectrum was obtained following \cite{PlanckCollaboration2016PlanckParameters}. %Note that, for TE and EE spectra, CamSpec provides co-added spectra.

The relation \eqref{eqn:Cl_equals_Glk_Pk} must hold for each of TT, TE and EE with the same primordial power spectrum on the same $k$ grid. Thus, we join these data sets together by creating a block matrix:
\begin{align}
    C_{L}^\mathrm{tot} = 
    \begin{pmatrix}
    \begin{array}{c} \def\arraystretch{2.2}
        C_{\ell_\mathrm{min}}^\mathrm{TT} \\
        \vdots  \\
        C_{\ell_\mathrm{max}}^\mathrm{TT} \\
        \\ \hline \\ 
        \gamma C_{\ell_\mathrm{min}}^\mathrm{TE} \\
        \vdots  \\
        \gamma C_{\ell_\mathrm{max}}^\mathrm{TE} \\
        \\ \hline \\
        \gamma^2 C_{\ell_\mathrm{min}}^\mathrm{EE} \\
        \vdots  \\
        \gamma^2 C_{\ell_\mathrm{max}}^\mathrm{EE} \\
    \end{array}
    \end{pmatrix}_L, \quad 
    G_{Lk}^\mathrm{tot} =
    \begin{pmatrix}
    \begin{array}{c} \def\arraystretch{2.2}
        \cdots G_{\ell_\mathrm{min} k}^\mathrm{TT} \cdots \\
        \vdots  \\
        \cdots G_{\ell_\mathrm{max} k}^\mathrm{TT} \cdots \\
        \\ \hline \\ 
        \cdots \gamma G_{\ell_\mathrm{min} k}^\mathrm{TE} \cdots \\
        \vdots  \\
        \cdots \gamma G_{\ell_\mathrm{max} k}^\mathrm{TE} \cdots \\
        \\ \hline \\
        \cdots \gamma^2 G_{\ell_\mathrm{min} k}^\mathrm{EE} \cdots \\
        \vdots  \\
        \cdots \gamma^2 G_{\ell_\mathrm{max} k}^\mathrm{EE} \cdots \\
    \end{array}
    \end{pmatrix}_{L k}.  \label{eqn:block_matrices}
\end{align}
Note that $\ell_\mathrm{min}$ and $\ell_\mathrm{max}$ are not the same across TT, TE and EE in practice, so the blocks have varying sizes. In terms of the new concatenated matrices, we have
\begin{align}
    C_L^\mathrm{tot} = \sum_k G_{Lk}^\mathrm{tot} P_k. \label{eqn:CL_tot_equals_GLk_Pk}
\end{align}
With these concatenated matrices, the MRL update for $P_k$ in \eqref{eqn:MRL_iteration} now includes contributions from each of the TT, TE and EE, with the latter two weighted by factors of $\gamma$ and $\gamma^2$. The covariance matrix appearing in $Q_L$ of the MRL iterations contains the covariance matrices for TT, TE and EE spectra along the diagonal, as well as the cross-covariances between them. This ensures that the $\tanh$ weights can still be significant if, for example, the $TT$ fit is good but the correlated TE and EE counterparts deviate significantly from the data. Note that cosmic variance limited values were used for the cross-covariance, as the full covariance matrix was unavailable during the development of this work. This is a crude approximation for estimating the joint likelihood, especially for high-$\ell$ correlations and polarization. However, in the context of MRL, this simply reduces the interaction between TT, TE, and EE contributions to the $\tanh$ weights in \eqref{eqn:MRL_iteration}, and has a small effect in the final reconstruction.

The parameter $\gamma$ in \eqref{eqn:block_matrices} is a tunable parameter which determines the relative weight of the polarization and temperature data within the MRL iterations. The covariance matrix for $C^\mathrm{tot}$ also contains appropriate factors of $\gamma$. Since the data vector and covariance matrix are rescaled by the same ($\ell$-dependent) factor, the $\chi^2$ statistic and the likelihood remain unchanged. However, this does affect the MRL reconstruction results because MRL is inherently non-linear. We keep $\gamma < 1$, motivated by the fact that the \textit{Planck} TT data has higher signal-to-noise than the TE and EE counterparts overall. This enters the MRL iterations through $\tilde{G}_{\ell k}$ and lowers the impact of fractional errors in $C_\ell^\mathrm{EE}$ and $C_\ell^\mathrm{EE}$ on updating $P_k$. We also found that the noisy EE data at high $\ell$s (often being negative) can sometimes cause numerical instabilities unless weighted down, since the fractional updates in \eqref{eqn:MRL_iteration} can struggle when $C_{\ell}^{(i)}$ crosses zero. In practice, we chose $\gamma=0.001$, the largest value where we could guarantee a stable reconstruction process. 

We measure the goodness of fit to the data at each iteration via the (reduced) chi-squared statistic:
\begin{align}
    \tilde{\chi}^2_{XY} \equiv \frac{1}{n_{XY}} \sum_{\ell, \ell'} \Delta C^{XY}_{\ell} \; \mathrm{Cov}^{-1}(\ell, \ell') \; \Delta C^{XY}_{\ell'},
\end{align}
where the residual $\Delta C_{\ell} \equiv C^\mathrm{obs}_{\ell} - C^\mathrm{recon}_{\ell}$ and $n_{XY}$ is the number of (unbinned) data points.

\subsection{Regularization}

The MRL algorithm provides a set of free-form PPS which improves the fit to the observed data compared to the power-law spectrum. This is a data-driven technique that can provide insights into the potential locations of primordial features. An important consideration is that, as the number of iterations increases, the algorithm may start to introduce sharp, rapid oscillations in the reconstruction that mainly fit the noise in the data. Regularizing the result is therefore crucial when searching for more physically viable forms of the primordial power spectrum.

In our previous works \cite{Hazra2013wmapMRL}, the reconstructed PPS obtained from the MRL algorithm was then smoothed by convolving it with a Gaussian kernel in the $\log$ space. On a discretised $k$ grid, this is given by
\begin{align}
    P^\mathrm{Smooth}_k = \frac{ \sum_{k'} P_{k'} \exp \left[ - \left( \frac{\log k' - \log k}{\Delta_\mathrm{smooth}} \right)^2 \right] } {\sum_{k'} \exp \left[ - \left( \frac{\log k' - \log k}{\Delta_\mathrm{smooth}} \right)^2 \right]
    },  \label{eqn:gaussin_smoothing}
\end{align}
for a tunable smoothing scale $\Delta_\mathrm{smooth}$. This method provides a smoother PPS that retains most of the features which improve the fit to data, and therefore mitigating the overfitting issue \cite{Hazra2013wmapMRL}. The smoothing scale can be further varied across the $k$ range to filter out features with different oscillation frequencies.

Incorporating the CMB polarization auto and cross spectra can act as a type of regularization by itself. Overfitting the noise in the CMB TT $C_\ell$s through large features in the PPS often results in a worse fit to the TE and EE counterparts. Despite this fact, we found that the PPS reconstructed using MRL or other methods still tends to sacrifice the goodness of fit to TE and EE spectra to improve the fit to temperature.

In this work, we focus on free-form PPS which can improve the fit to \textit{each} of the TT, TE and EE spectra. One of the main challenges for achieving this via MRL was that the algorithm prioritises fitting the TT data over TE and EE with smaller signal-to-noises overall. We found that smoothing the reconstructed spectrum using \eqref{eqn:gaussin_smoothing} allows us to counter this effect. However, this smoothing method requires a choice of the kernel which is, to some extent, arbitrary. The smoothing is also applied after the MRL iterations and is oblivious to the importance of oscillatory features in fitting the data. This way of smoothing the reconstructed primordial spectrum depends on the value of $\Delta$, the width of smoothing, and also treat the whole range of the spectrum uniformly even though different forms of features may exist at different wavenumbers.

To address this issue and to further improve the regularization process, we developed a novel approach by directly including a smoothing term at each MRL iteration. Inspired by the diffusion (heat) equation
\begin{align}
    \frac{\partial f}{\partial t} = \kappa \frac{\partial^2 f}{\partial x^2},
\end{align}
which causes the peaks and troughs of $f(t)$ to diffuse away over time, we modify the MRL iteration step. The regularized MRL (RegMRL) algorithm is given as follows.

\begin{align}
	&P^{(i+1)}_k = P^{(i)}_k [ 1 + \sum_\ell  \frac{C^\mathrm{obs}_\ell - C^{(i)}_\ell}{C^{(i)}_\ell} \tilde{G}_{\ell k} \nonumber  \tanh^2 \left[ Q^{(i)}_\ell \left( C^\mathrm{obs}_\ell - C^{(i)}_\ell \right)  \right] + \kappa \sum_{k'} (D^2)_{k k'} P^{(i)}_{k'} ]. \label{eqn:regMRL_iteration}
\end{align}

The matrix, $D$, is a discrete analogue of the differential operator $\partial/\partial k$ \footnote{When discretising the second order differential operator $\partial^2/\partial k^2$ to obtain $D^2$, we use forward differences for the inner derivative and backward differences for the outer derivative.}. At each iteration, the MRL term in \eqref{eqn:regMRL_iteration} pushes $P(k)$ towards a direction which improves the overall fit to data, while the regularization term led by $\kappa$ acts to `diffuse' the sharp fluctuations present in the spectrum. Therefore, $\kappa$ here acts as a regularization parameter: the larger the $\kappa$, the smoother the reconstruction. We retrieve the usual MRL when $\kappa=0$. Note that we have used a multiplicative update (with an extra factor of $P^{(i)}_k$) for the diffusion term in \eqref{eqn:regMRL_iteration} to be consistent with RL iterations.

We note that, while we are the first to include this regularization term within the MRL iterations, this method of regularization itself is far from new. Diffusion-based techniques have been used in various fields as a means to regularize ill-posed inverse problems. The term $\sum_{k'}D_{kk'}P_{k'}$ may also appear in gradient descent updates of an optimisation problem if the loss function contains a $L_2$ roughness penalty of the derivative $P'(k)$. \footnote{The penalty term is proportional to $\int dk |P'(k)|^2$. After discretisation, this is equivalent to a Tikhonov regularization of $\| \Gamma \mathbf{P} \|^2$ with $\Gamma$ being the discrete version of the differential operator $\partial/\partial k$.} In fact, similar regularization techniques for the RL (not MRL) algorithm have been studied in \cite{Katsaggelos1991AAlgorithm,Conchello1996FastMicroscopy} for image analysis applications.

We also consider another popular regularization method for the RL algorithm in the field of image analysis: total variation (TV) regularization. This involves a penalty term proportional to the integral of the gradient magnitude. In one dimension, the penalty term is ${\int dx\; \lvert\partial f / \partial x\rvert}$. This term is not differentiable but can be minimised numerically; we replace the diffusion term in \eqref{eqn:regMRL_iteration} by
\begin{align}
    \kappa \sum_{k'} D_{k k'} \; \mathrm{sign}(\sum_{k''} D_{k' k''} P^{(i)}_{k''}),
\end{align}
where $\mathrm{sign}(x)=-1,0,1$ for negative, zero and positive $x$, respectively. We will refer to this iterative method as MRL-TV. TV regularization eventually forces the function (or image) to have as many vanishing derivatives as possible, similar to regression methods with $L_1$ penalty terms such as LASSO regression.

\section{\label{sec:results} Results and Discussion}

\subsection{Image deblurring applications}

In this work, we developed two methods named RegMRL and MRL-TV for a robust reconstruction of the primordial power spectrum (PPS) from the CMB data. These methods, described in detail in the previous section, can also be applied to deblurring (deconvolution) tasks of various images. The corresponding inverse problem can be written as ${g(i) = \sum_j p(i,j) f(j)}$ (cf. \eqref{eqn:Cl_equals_Glk_Pk}), where the original image $f(i)$ is convolved with a point spread function (PSF) $p(i,j)$ to give the observed image $g(i)$. The index $i$ denotes the pixel number.

\begin{figure}
\centering
\includegraphics[width=0.8\textwidth]{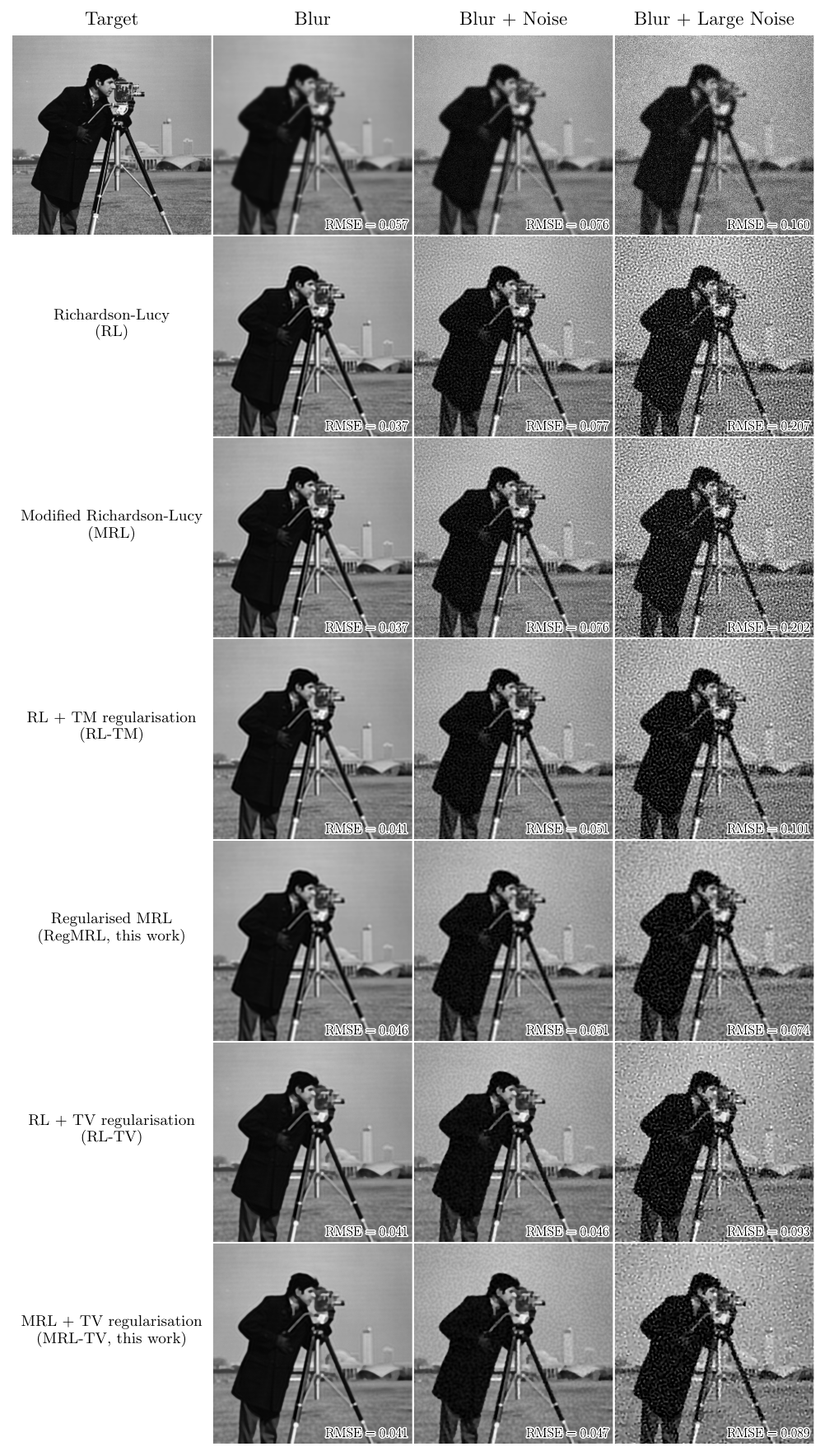}
\caption{\label{fig:image_analysis} The deconvolution algorithms discussed in this paper are applied to a test image (cameraman), blurred with a Gaussian kernel and added additional Gaussian noise. Both RL and MRL accurately recover most of the features present in the original image but perform worse for noisy images due to noise amplification. Methods with regularisation yield better reconstructions with less noise and better image quality. The root mean squared errors (RMSE) between the reconstruction and the target (original) image are shown.}
\end{figure}

We test the algorithms on a standard test image and show the results in Figure \ref{fig:image_analysis}. The original image is blurred using a Gaussian kernel with a full width at half maximum (FWHM) of $\approx20$ pixels. We then add some Gaussian white noise to the blurred image to simulate a noisy imaging system; the three test images have noise standard deviations equal to 0, 0.05, and 0.15 in units where black is 0 and white is 1. For a quantitative analysis, we compute the root mean squared error (RMSE) of the reconstructed image after 50 iterations with respect to the true image. For comparison, we also implemented the conventional regularized Richardson-Lucy algorithms from image analysis: Tikhonov-Miller (RL-TM) and total variation (RL-TV), analogous to RegMRL and MRL-TV, respectively. We see that our results are competitive with these methods and potentially can have advantages in some cases.

We observe that the methods without regularization (RL and MRL) yield marginally better results for the noiseless image but perform badly with noise present. They can accurately recover the sharp features in the image while amplifying the noise in the process. Both regularization methods, RegMRL and MRL-TV, drastically reduce the error in the reconstruction of noisy images: RMSE is reduced by a factor of 2.7 and 2.3, respectively. The resulting reconstruction still retains most of the features present in the image. regularization is therefore crucial for scientific analyses where the quantitative error matters \footnote{This may not always be the case in some image analysis applications for which having clear edges and/or pleasant-looking images matters.}. 

\subsection{Validation using simulated data}

\begin{figure}[htbp!]
\includegraphics[width=0.9\textwidth]{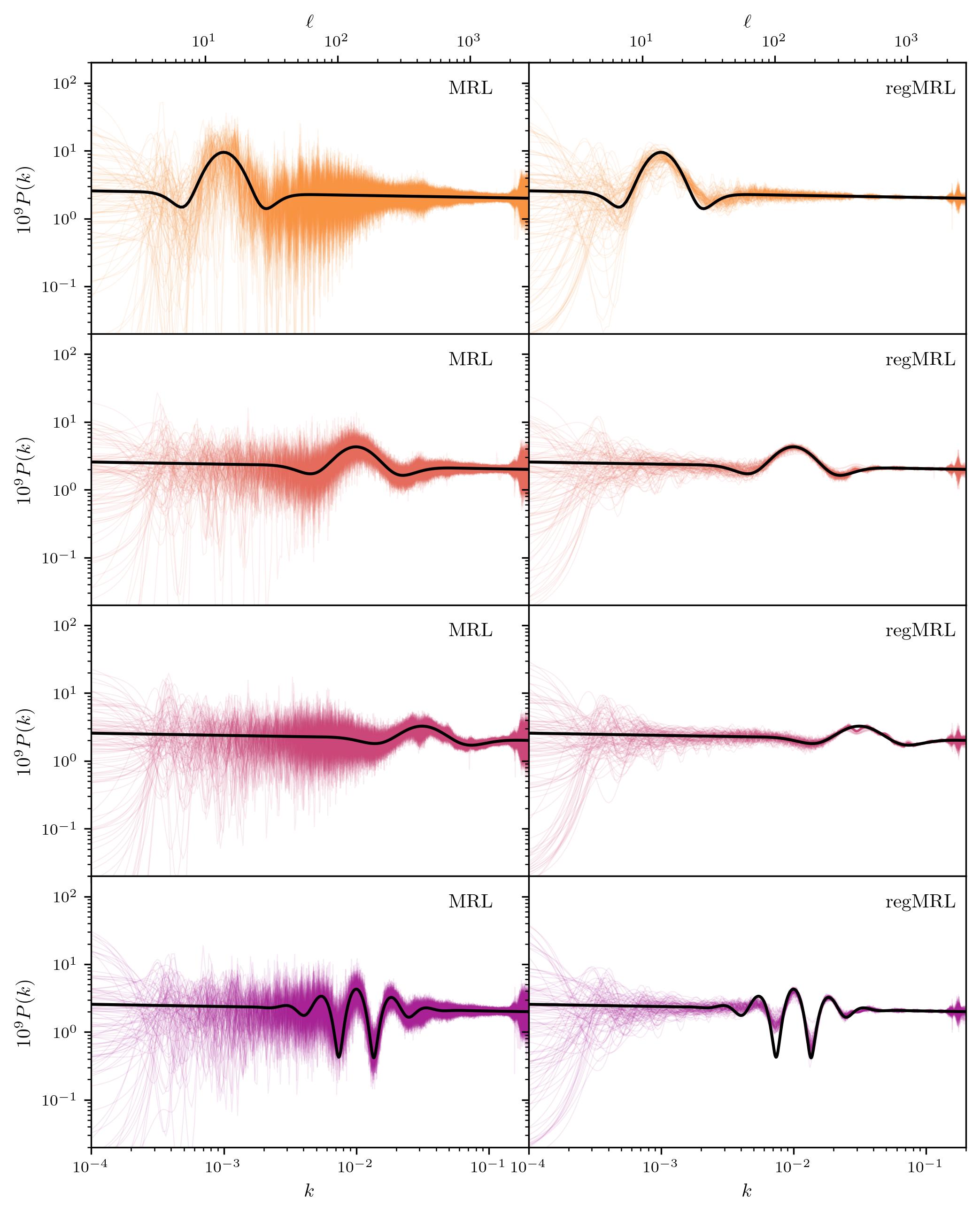}
\caption{\label{fig:retrieval_test} Recovery of various primordial power spectra from mock data using MRL (left) and RegMRL (right). Based on 4 different forms of power spectra with artificial features shown above (black), we compute the $C_\ell$'s and draw 100 realisations from them using full covariance matrices (see text for details). The reconstructed $P(k)$'s for each of these realisations are drawn in thin coloured lines. MRL successfully retrieves the underlying features but has relatively large fluctuation due to noise amplification. The regularization introduced in this work significantly reduces the noise and allows us to recover the features more clearly. Note that previous studies on MRL \cite{Hazra2013wmapMRL} did not use the noisy power spectra shown on the left; they were smoothed post-reconstruction using a Gaussian kernel. The $\ell$ axis is computed using Limber approximation as a rough guide to the scales; $l\approx k \eta_\text{rec}$.}
\end{figure}

We validate the reconstruction method on some mock data. We create several forms of PPS with artificially added features, compute the corresponding theoretical $C_\ell$s, and draw 100 random realisations from them. The $C_\ell^\mathrm{TT}$ for $\ell<30$ were drawn from a $\chi^2$-distribution (up to a constant factor), whereas the rest were drawn from a multivariate normal distribution using the CamSpec unbinned TT, TE and EE covariance matrices. We assume cosmic variance limited cross covariances (TT-TE, TE-EE, and TT-EE). Figure \ref{fig:retrieval_test} shows the result of MRL and RegMRL reconstructions applied to these mock $C_\ell$s to recover the PPS. 

We find that MRL can indeed recover the features present in PPS as it has been shown in \cite{Shafieloo2004MRL,Hazra2013wmapMRL,Hazra2014PlanckMRL}. Note that, in these previous works, the reconstruction results were smoothed using a Gaussian smoothing kernel to regularize them. Here, we use the diffusive regularization method outlined in the methods section and obtain the results shown on the right-hand side of Figure \ref{fig:retrieval_test}. The regularization enables us to retrieve the features more clearly.

We note that for the bottom two plots of Figure \ref{fig:retrieval_test}, the RegMRL reconstruction results do not perfectly trace the true underlying $P_\zeta(k)$'s. The reason is twofold. First, the diffusive smoothing term tends to smooth out high-frequency oscillations which do not contribute significantly to fitting the data. For example, RegMRL reconstructs the dip around $k=7\times10^{-3}$ to be smaller than the true value. Second, the inherent degeneracy of the inverse problem may direct the algorithm to converge on another form of PPS which gives nearly identical $C_\ell$s. The wavenumbers ($k$) and the multipoles ($\ell$) have a one-to-one correspondence under the Limber approximation ($l\approx k \eta_\text{rec}$\footnote{Here, $\eta_\text{rec}$ denotes conformal distance to the last scattering surface.}), but in reality, each $k$ contributes to multiple $\ell$s that are near or smaller than $k\eta_\text{rec}$ (larger scales). Excess powers can therefore compensate for a lack of power on nearby scales. The third row of Figure \ref{fig:retrieval_test} shows one such result. Both MRL and RegMRL results have a small dip around $k=2\times10^{-2}$, compensated by small oscillations around the true $P(k)$ for before and after. The corresponding $C_\ell^\mathrm{TT}$s are within the fractional error of $O(10^{-3})$ with respect to the true values. Extra fluctuation around the true feature are also small enough to avoid suppression from diffusion.

\subsection{Reconstruction of PPS from \textit{Planck} CMB}

\begin{figure}[htbp!]
\centering
\includegraphics[width=0.95\textwidth]{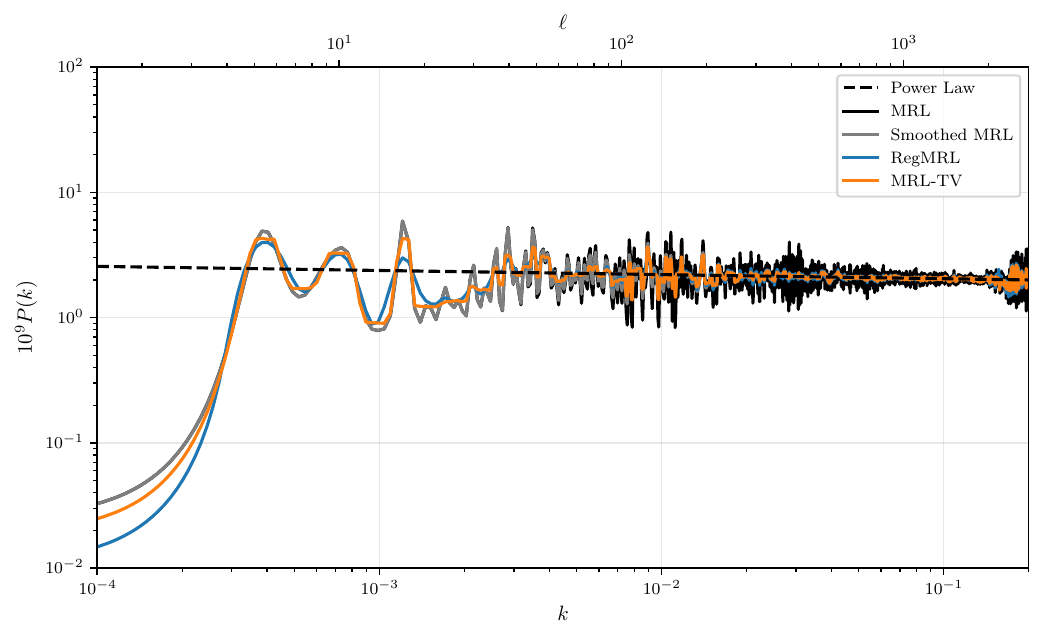}
\caption{\label{fig:regMRL_versus_smoothing_one} Primordial power spectrum reconstructed from the CamSpec TTTEEE data using MRL (black), MRL smoothed using a Gaussian kernel with $\Delta=0.03$ (grey), the RegMRL with $\kappa=0.01$ (blue), and MRL-TV with $\kappa = 0.008$ (orange) after 50 iterations. We observe some common features within all three spectra but their respective amplitudes vary due to the differences in the regularization methods. Note the log scaling in both axes; the near-scale-invariant power spectrum appears as a straight line (black dashed). The $\ell$ axis is computed using Limber approximation as before.}
\end{figure}

We now apply the method outlined in the previous section to the CamSpec data specified in the methods section. We show the reconstructed results from MRL, MRL followed by Gaussian kernel smoothing, RegMRL, and MRL-TV in Figure~ \ref{fig:regMRL_versus_smoothing_one}. The features in the reconstructed power spectra can be categorized in three parts: suppression and dip at the largest scales ($k<10^{-3}/{\rm Mpc}$), broad features at the intermediate scales ($k<10^{-3}-10^{-2}/{\rm Mpc}$) and a combination of oscillatory features at different frequencies at the small scales ($k<10^{-2}-10^{-1}/{\rm Mpc}$). The features at the largest and intermediate scales were present since WMAP with similar significance. It has been shown that PPS reconstructed from \textit{Planck} data at different MRL iterations and smoothing scales also provide a better fit to WMAP data, compared to the power-law spectrum \cite{Hazra2014PlanckMRL}.

Several inflationary models have been proposed that effectively fit these `outliers'. An initial period of moderate fast-roll~\cite{Hazra2014InflationSpectra}, punctuated inflation~\cite{Jain2009PunctuatedMultipoles} or kinetic dominated initial conditions~\cite{Contaldi2003SuppressingAnisotropies,Hergt2019CaseInflation,Ragavendra2022SuppressionBispectra} generate a suppression in power at the largest scales. A transition in the scalar field potential~\cite{Starobinsky1992SpectrumPotential} or the presence of a step~\cite{Adams2001InflationaryStep,Covi2006InflationFuture,Hazra2010PrimordialPotential,Chluba2015,Miranda2016NonlinearSpectra} generates a dip and a bump which can fit the $\ell\sim22$ outlier. These types of models may also produce sinusoidal oscillations or sharp features in the power spectrum. Wiggly Whipped Inflation (WWI) combines these two types of features and finds several candidates as local best fits to the \textit{Planck} data~\cite{Hazra2014WigglyInflation,Hazra2016PrimordialPolarization,Hazra2018ProbingObservations,Hazra2021InflationBeyond}. Axion monodromy models~\cite{Chen2008GenerationInflation,Flauger2010OscillationsInflation,Flauger2011ResonantNon-Gaussianity,Aich2013OscillationsDatasets} generate logarithmically spaced oscillations that fit possible outliers at intermediate to small scales. The primordial power spectrum from the Classical Primordial Standard Clock (CPSC) models~\cite{Chen2011FingerprintsPerturbations,Chen2012PrimordialInflation,Chen2015ModelsClock,Braglia2021ComparingData,Braglia2022PrimordialAnomalies,Braglia2022UncoveringSpectra} contains a combination of sharp and resonant features that provide a global best fit to the \textit{Planck} data. The dip around $\ell=750$ is effectively addressed by WWI and CPSC models. Features from these two models help in fitting temperature and polarization outliers jointly. Recently, two of the authors of this paper have co-authored the development of One Spectrum~\cite{Hazra2022OneCosmology} and the corresponding theoretical model building~\cite{Antony2022DiscordancesDynamics} with Hubble flow functions that generate features at small scales and also helps to resolve tensions between CMB and low-redshift observations.

\begin{figure}[htbp!]
\centering
\includegraphics[width=\textwidth]{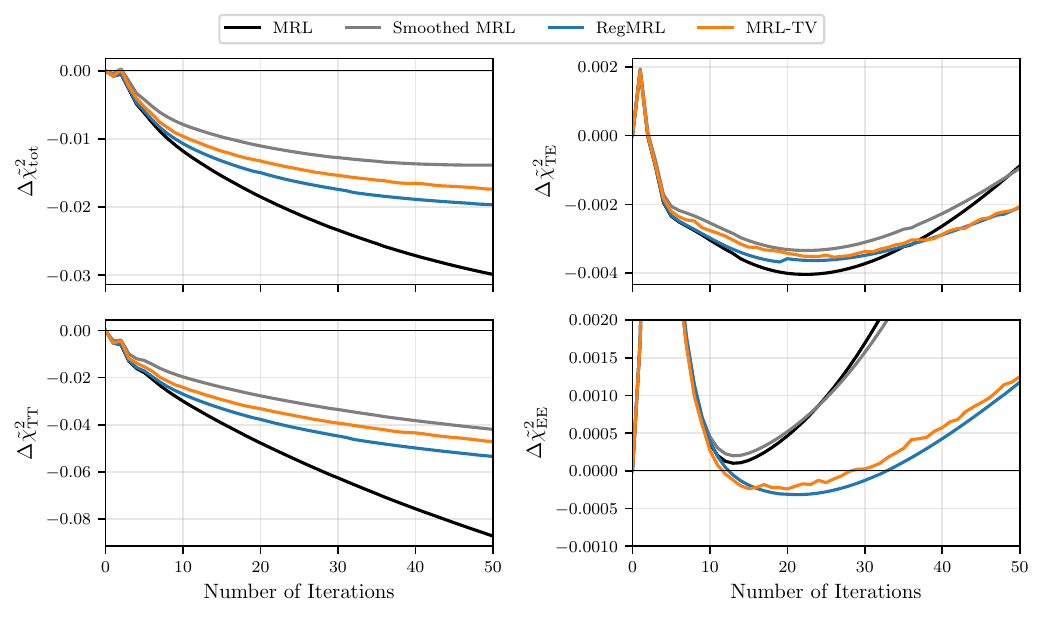}
\caption{\label{fig:chisquare_versus_iterations} Improvements in the chi-squared fit to the CamSpec data for MRL, smoothed MRL, RegMRL and MRL-TV reconstructions discussed in the main text. The four plots show normalised chi-squared improvements to the combined, TT, TE, and EE data. All three methods improve the fit to individual spectrum during early iterations, but later they sacrifice the fit to TE and EE $C_\ell$s in favour of the TT counterpart. MRL (black) without smoothing or regularization gives the best improvement to the overall fit, albeit with many rapid oscillations in the spectrum (Figure~ \ref{fig:regMRL_versus_smoothing_one}) that are less likely to be physical. Smoothing MRL using a Gaussian kernel (grey) removes these oscillations whilst worsening the fit to data. RegMRL (blue) and MRL-TV (orange) achieve a better fit to the data, especially on the EE data.}
\end{figure}

Figure \ref{fig:chisquare_versus_iterations} shows the $\tilde{\chi}^2$ values obtained for the reconstructed PPS at each iteration, obtained using the four methods shown in Figure \ref{fig:regMRL_versus_smoothing_one}. The four plots indicate the improvement in $\tilde{\chi}^2$ with respect to the power-law PPS for the combined, TT, TE, and EE datasets. We observe that all three methods improve the fit to each of the TT, TE, and EE $C_\ell$s during early iterations. With a large number of iterations, however, all methods start to sacrifice the fit to the polarization (TE and EE) $C_\ell$s in favour of the temperature (TT) counterpart. MRL reconstructions, for example, always have a worse fit to the polarization data than the power-law spectrum. The algorithm then keeps on `overfitting' the noise in the TT data in order to improve the total $\tilde{\chi}^2$ further.

Smoothing MRL using a Gaussian kernel removes many rapid oscillations present in the spectrum obtained from MRL (Figure~ \ref{fig:regMRL_versus_smoothing_one}) that are less likely to be physical. However, smoothing also increases $\tilde{\chi}^2$. This is partially because the smoothing is applied to soften the features present regardless of how much they affect $\tilde{\chi}^2$. RegMRL addresses this issue by having the regularization term within each iteration; only the features with significant improvements in the fit can remain. As can be seen from Figure \ref{fig:chisquare_versus_iterations}, RegMRL and MRL-TV give better overall $\Delta\tilde{\chi}^2$ compared to smoothed MRL. The fit to EE data, in particular, benefits significantly from these new regularization techniques. We see that regularization is necessary for us to obtain any improvement ($\Delta\tilde{\chi}^2<0$) for the EE data.

We note that there are points in the iterations where the fits to each of TT, TE, and EE $C_\ell$s are improved simultaneously. The overall fit to polarization data is optimised around iteration 22, where RegMRL gives $\Delta \tilde{\chi}^2_{TT} = -3.94\times 10^{-2}$, $\Delta \tilde{\chi}^2_{TE} = -3.62\times 10^{-3}$, and $\Delta \tilde{\chi}^2_{EE} = -3.13\times 10^{-4}$. MRL-TV also yields similar improvements; $\Delta \tilde{\chi}^2_{TT} = -3.46\times 10^{-2}$, $\Delta \tilde{\chi}^2_{TE} = -3.50\times 10^{-3}$, and $\Delta \tilde{\chi}^2_{EE} = -1.71\times 10^{-4}$. 

\begin{figure}[htbp!]
\includegraphics[width=\textwidth]{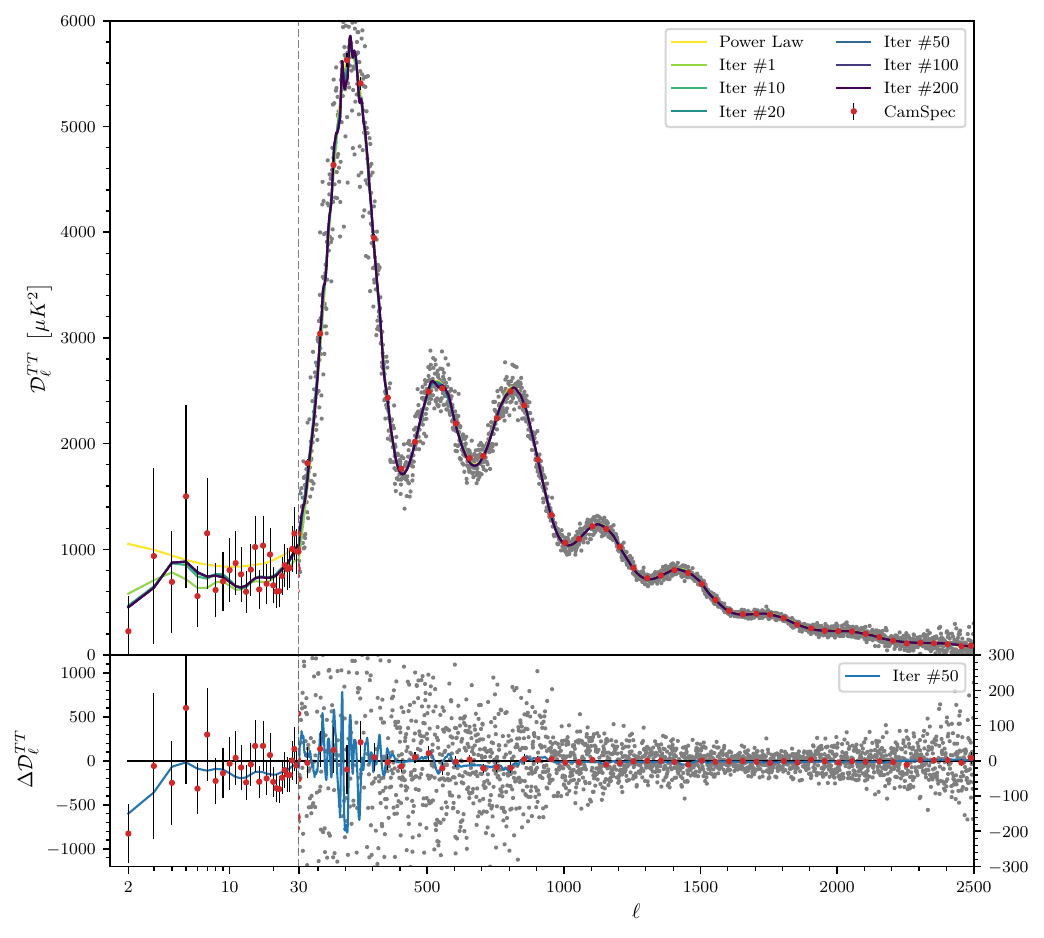}
\caption{\label{fig:rec_cl_TT} Temperature angular power spectra for the reconstructed primordial power spectra from RegMRL with CamSpec unbinned data taken from \cite{Efstathiou2021AMaps}. Transfer functions for the projection are computed using CAMB \cite{Lewis2000camb} using the best-fit parameters to CamSpec TTTEEE. Shown below is the residual with respect to the $\Lambda$CDM best-fit $C_\ell$s. Note the transition from log scale to linear at $\ell=30$ in the plot.}
\end{figure}

In Figure \ref{fig:rec_cl_TT}, we show $C_\ell^\mathrm{TT}$s computed from the MRL reconstructions at different iterations, together with its residuals.Note that $D_\ell^{XY} = \ell(\ell+1) C_\ell^{XY}/2\pi$. The unbinned $C_\ell$ data points are shown in grey, while the binned $C_\ell$s with $\Delta\ell=50$ are plotted in red together with their error bars.

For $\ell\lesssim 10$, we find that the MRL reconstructions tend to fit the low observed $C_\ell^\mathrm{TT}$s and yield values lower than the $\Lambda$CDM best fit. In particular, the low quadrupole ($\ell=2$) in the data drives the low-$k$ PPS values towards zero, as can be seen in Figure ~\ref{fig:regMRL_versus_smoothing_one}. Due to the large cosmic variances present in this multipole range, the statistical significance of these outliers needs to be assessed with care. Note that if the value of $C_2$ is set to be larger than the best-fit value, then the corresponding MRL reconstructions would no longer be suppressed at low-$k$ as seen in Figure~\ref{fig:regMRL_versus_smoothing_one}. 

The dip around $\ell=22$, in particular, has been found in the WMAP first-year data release and is also present in \textit{Planck} \cite{PlanckCollaboration2013likelihood}. \textit{Planck} has estimated the tension between the CMB power spectrum in $\ell\lesssim 40$ and best-fit $\Lambda$CDM model to be 2.5$\sigma$-3.0$\sigma$ \cite{PlanckCollaboration2013likelihood}. The reconstructed spectra show a dip around $k\sim2\times10^{-3}~{\rm Mpc}^{-1}$. If some new physics is responsible for the outlier, it is likely present in the EE spectrum. However, given the low signal-to-noise ratio at the largest scale E-mode observation in Planck, we have yet to verify this feature.

For $\ell\ge 30$, the MRL reconstruction exhibits larger fluctuation as the number of iterations increases. These added fluctuation lie within $1\sigma$ of the binned $C_\ell$ data points. The largest deviations from the best fit $C_\ell$s are located in the range $30\le\ell\le 500$ and near $\ell=750$. These roughly correspond to $2\times 10^{-3} \le k \le 5\times 10^{-2}/\mathrm{Mpc}$ and $6\times10^{-2}/\mathrm{Mpc}$ in $P(k)$ where large features are visible in the reconstructions (Figure~ \ref{fig:regMRL_versus_smoothing_one}).

\subsection{Testing statistical significance}

\begin{figure}
\centering
\includegraphics[width=0.8\textwidth]{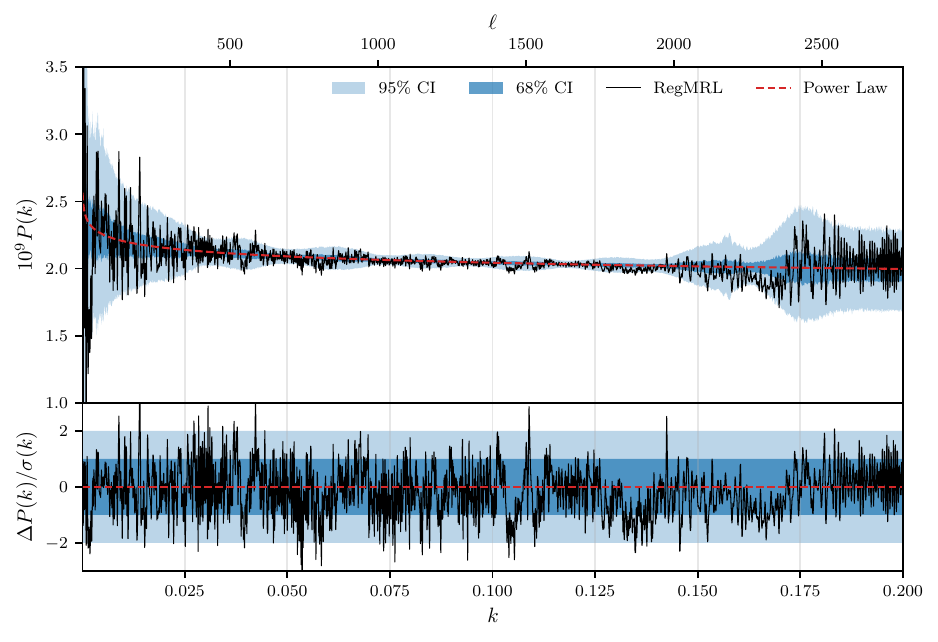}
\\
\includegraphics[width=0.8\textwidth]{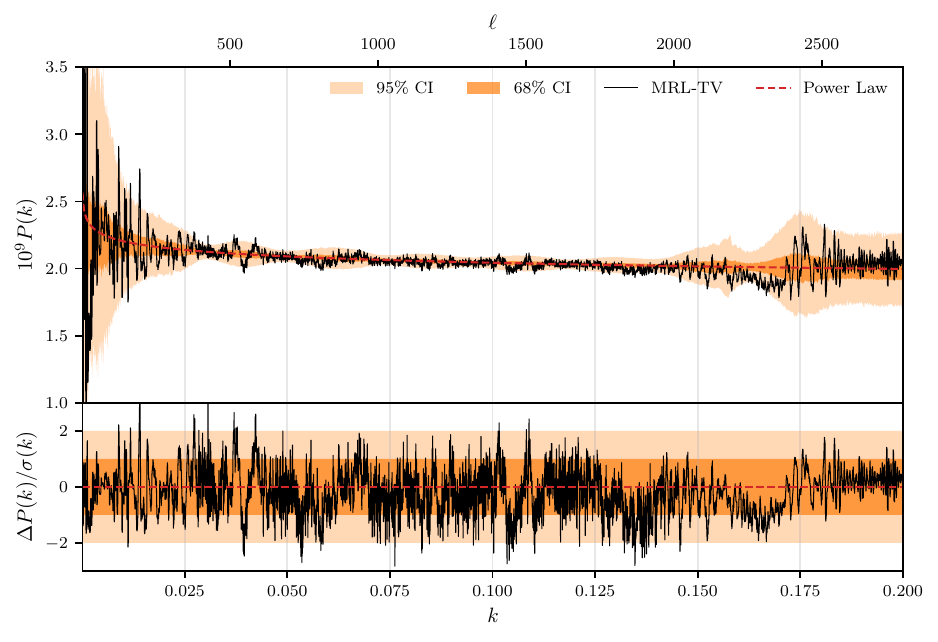}
\caption{\label{fig:gaussian_sims}
Primordial power spectrum reconstructed from the CamSpec TTTEEE data after 22 iterations (when $\tilde{\chi}^2_\mathrm{EE}$ is minimised) using RegMRL (top) with $\kappa=0.01$ and MRL-TV (bottom) with $\kappa=0.008$. We apply the same algorithm to reconstruct $P(k)$ from 1000 sets of mock $C_\ell$s based on the best-fit power-law PPS and provide the 68\% and 95\% confidence intervals at each $k$. Below, we plot the reconstructed spectrum's deviations from the median value obtained from simulated values, divided by the standard deviation $\sigma(k)$ at $k$, also obtained from simulations. The $\ell$ axis is computed using Limber approximation as a rough guide to the scales. The increased values of $\sigma(k)$ at high $k$s are due to the lack of polarization data at $\ell>2000$ and temperature at $\ell>2500$. Overall, we do not find any statistically significant deviations from the power-law spectrum.
}
\end{figure}

We performed a simulation-based analysis in order to assess the statistical significance of the features present in the reconstructed spectra. RegMRL and MRL-TV were applied to 1000 sets of simulated $C_\ell^\mathrm{TT}$, $C_\ell^\mathrm{TE}$ and $C_\ell^\mathrm{EE}$s, generated based on the power-law PPS. Here, $C_\ell$s were drawn similarly to the ones used for Figure \ref{fig:retrieval_test}; $C_\ell$s under $\ell<30$ were drawn from a chi-squared distribution with $2\ell+1$ degrees of freedom, while, for $\ell>30$, $C_\ell$s were drawn from a multivariate normal distribution using the full CamSpec covariance matrix. Cross-covariances between datasets (TT, TE and EE) were set to the cosmic variance values. Figure \ref{fig:gaussian_sims} shows the results.

We test the null hypothesis $H_0$: `data is consistent with a power-law PPS' against the alternative hypothesis $H_1$: `features in the PPS are required to explain the CMB data'.\footnote{Note that these hypotheses assume that the other cosmological parameters are fixed to the best-fit values used in the analysis.} We form our $p$-value statistic at each $k$ by comparing the features against the distribution of PPS values obtained from the power-law simulations. The 68\% and 95\% confidence intervals (CIs), approximately equal to the $1\sigma$ and $2\sigma$ bounds, are evaluated from the 1000 simulations. As can be seen in Figure \ref{fig:gaussian_sims}, we have few features that lie outside the 95\% CIs and none outside the 99\% CIs, and therefore do \textit{not} reject the null hypothesis; the data is consistent with having no features.

Note that we treated each $k$ independently in this analysis without adjusting for the `look-elsewhere effect' caused by the large number of $k$ points involved. This is, in general, more favourable for detecting features since only one of many $p$-values (for each $k$) needs to be significant. An adjusted statistic is required to incorporate the number of effectively independent tests that are being made and the probability of having at least one large $p$-value. In this analysis, however, this is not as crucial because we do not detect any statistically significant result even without adjusting for the look-elsewhere effect.

\section{\label{sec:conclusion} Conclusion}

In this work, we produced free-form reconstructions of the primordial power spectrum (PPS) based on the CMB temperature and polarization data from Planck. The MRL algorithm has been used to iteratively reconstruct the PPS from unbinned CamSpec $C_\ell$ data and the full covariance matrix. We introduced two new regularization approaches inspired by the diffusion equation and total-variation regularization.

We focused on forms of the PPS that can improve the fit to the individual $C^\mathrm{TT}_\ell$, $C^\mathrm{TE}_\ell$ and $C^\mathrm{EE}_\ell$ \textit{simultaneously}, and not simply by boosting the fit to one at the cost of worsening another. Such PPS were found using the regularized MRL methods introduced in this work. It is also worth noting that the reconstructed result retains many of the features found in the previous temperature-only analyses.

We studied the statistical significance of the features found in the reconstructed PPS by comparing them with those from 1000 mock $C_\ell$s drawn from PPS without features. Overall, we did not reject the null hypothesis and the data is consistent with a featureless power-law spectrum.

Free-form reconstruction techniques studied in this work serve as useful tools for the search for features in the PPS. The iterative approach developed in this work is complementary to other methods such as minimally parametric approaches and direct template fitting. Furthermore, as the methodology was extended to include CMB polarization, it will benefit further from the forthcoming CMB surveys with increased polarization sensitivities. We plan on extending our analysis to other existing and future CMB experiments.

\section*{Acknowledgments}

We thank Satadru Bag and Rodrigo Calderon for meaningful discussions. A.S. would like to acknowledge the support by the National Research Foundation of Korea (NRF-2021M3F7A1082053) and the support of the Korea Institute for Advanced Study (KIAS) grant funded by the government of Korea.


\begin{thebibliography}{99}


\bibitem{Martin2014EncyclopdiaInflationaris} Martin, J., Ringeval, C. \& Vennin, V. Encyclopædia Inflationaris. {\em Physics Of The Dark Universe}. \textbf{5-6} pp. 75-235 (2014,12), https://linkinghub.elsevier.com/retrieve/pii/S2212686414000053

\bibitem{Meerburg2019PrimordialNon-Gaussianity} Meerburg, P., Green, D., Abidi, M., Amin, M., Adshead, P., Ahmed, Z., Alonso, D., Ansarinejad, B., Armstrong, R., Avila, S. \& Others Primordial Non-Gaussianity. (arXiv,2019,3), https://arxiv.org/abs/1903.04409

\bibitem{Achucarro2022Inflation:Observations} Achúcarro, A., Biagetti, M., Braglia, M., Cabass, G., Caldwell, R., Castorina, E., Chen, X., Coulton, W., Flauger, R., Fumagalli, J. \& Others Inflation: Theory and Observations.  (2022,3), http://arxiv.org/abs/2203.08128

\bibitem{PlanckCollaboration2018parameters} Planck Collaboration Planck 2018 results. VI. Cosmological parameters. {\em Astronomy \& Astrophysics}. \textbf{641} pp. A6 (2020,9), https://www.aanda.org/10.1051/0004-6361/201833910

\bibitem{PlanckCollaboration2018inflation} Planck Collaboration Planck 2018 results. X. Constraints on inflation. {\em Astronomy \& Astrophysics}. \textbf{641} pp. A10 (2020)

\bibitem{Starobinsky1992SpectrumPotential} Starobinsky, A. Spectrum of adiabatic perturbations in the universe when there are singularities in the inflationary potential. {\em JETP Lett.}. \textbf{55} pp. 489-494 (1992)

\bibitem{Adams2001InflationaryStep} Adams, J., Cresswell, B. \& Easther, R. Inflationary perturbations from a potential with a step. {\em Physical Review D}. \textbf{64}, 123514 (2001,11), https://link.aps.org/doi/10.1103/PhysRevD.64.123514

\bibitem{Covi2006InflationFuture} Covi, L., Hamann, J., Melchiorri, A., Slosar, A. \& Sorbera, I. Inflation and WMAP three year data: Features have a Future!. {\em Phys. Rev. D}. \textbf{74} pp. 83509 (2006,6), http://arxiv.org/abs/astro-ph/0606452%20http://dx.doi.org/10.1103/PhysRevD.74.083509

\bibitem{Chen2008GenerationInflation} Chen, X., Easther, R. \& Lim, E. Generation and characterization of large non-Gaussianities in single field inflation. {\em Journal Of Cosmology And Astroparticle Physics}. \textbf{2008}, 010 (2008,4), https://iopscience.iop.org/article/10.1088/1475-7516/2008/04/010

\bibitem{Flauger2010OscillationsInflation} Flauger, R., McAllister, L., Pajer, E., Westphal, A. \& Xu, G. Oscillations in the CMB from Axion Monodromy Inflation. {\em Journal Of Cosmology And Astroparticle Physics}. \textbf{2010}, 009 (2010)

\bibitem{Flauger2011ResonantNon-Gaussianity} Flauger, R. \& Pajer, E. Resonant Non-Gaussianity. {\em Journal Of Cosmology And Astroparticle Physics}. \textbf{2011}, 017 (2011)

\bibitem{Aich2013OscillationsDatasets} Aich, M., Hazra, D., Sriramkumar, L. \& Souradeep, T. Oscillations in the inflaton potential: Complete numerical treatment and comparison with the recent and forthcoming CMB datasets. {\em Physical Review D}. \textbf{87}, 083526 (2013,4), https://link.aps.org/doi/10.1103/PhysRevD.87.083526

\bibitem{Chen2010} Chen, X. Primordial non-gaussianities from inflation models. {\em Advances In Astronomy}. \textbf{2010} (2010)

\bibitem{Slosar2019ScratchesFluctuations} Slosar, A., Chen, X., Dvorkin, C., Green, D., Meerburg, P., Silverstein, E. \& Wallisch, B. Scratches from the Past: Inflationary Archaeology through Features in the Power Spectrum of Primordial Fluctuations.  (2019,3), http://arxiv.org/abs/1903.09883

\bibitem{Hazra2022OneCosmology} Hazra, D., Antony, A. \& Shafieloo, A. One spectrum to cure them all: signature from early Universe solves major anomalies and tensions in cosmology. {\em Journal Of Cosmology And Astroparticle Physics}. \textbf{2022}, 063 (2022,8), https://iopscience.iop.org/article/10.1088/1475-7516/2022/08/063

\bibitem{DiValentino2021InSolutions} Di Valentino, E., Mena, O., Pan, S., Visinelli, L., Yang, W., Melchiorri, A., Mota, D., Riess, A. \& Silk, J. In the realm of the Hubble tension—a review of solutions. {\em Classical And Quantum Gravity}. \textbf{38}, 153001 (2021,7), https://iopscience.iop.org/article/10.1088/1361-6382/ac086d

\bibitem{Abdalla2022CosmologyAnomalies} Abdalla, E., Abellán, G., Aboubrahim, A., Agnello, A., Akarsu, Ö., Akrami, Y., Alestas, G., Aloni, D., Amendola, L., Anchordoqui, L. \& Others Cosmology intertwined: A review of the particle physics, astrophysics, and cosmology associated with the cosmological tensions and anomalies. {\em Journal Of High Energy Astrophysics}. \textbf{34} pp. 49-211 (2022), https://www.sciencedirect.com/science/article/pii/S2214404822000179

\bibitem{PlanckCollaboration2013inflation} Planck Collaboration Planck 2013 results. XXII. Constraints on inflation. {\em Astronomy \& Astrophysics}. \textbf{571} pp. A22 (2014,11)

\bibitem{PlanckCollaboration2015inflation} Planck Collaboration Planck 2015 results. XX. Constraints on inflation. {\em Astronomy \& Astrophysics}. \textbf{594} pp. A20 (2016)

\bibitem{Ichiki2010CosmicSpectrum} Ichiki, K., Nagata, R. \& Yokoyama, J. Cosmic discordance: Detection of a modulation in the primordial fluctuation spectrum. {\em Physical Review D}. \textbf{81}, 083010 (2010,4), https://link.aps.org/doi/10.1103/PhysRevD.81.083010

\bibitem{Meerburg2012WMAP7Spectrum} Meerburg, P., Wijers, R. \& Schaar, J. WMAP7 constraints on oscillations in the primordial power spectrum. {\em Monthly Notices Of The Royal Astronomical Society}. \textbf{421}, 369-380 (2012)

\bibitem{Meerburg2014SearchingApproach} Meerburg, P., Spergel, D. \& Wandelt, B. Searching for oscillations in the primordial power spectrum. I. Perturbative approach. {\em Physical Review D}. \textbf{89}, 19-26 (2014)

\bibitem{Meerburg2014SearchingData} Meerburg, P., Spergel, D. \& Wandelt, B. Searching for oscillations in the primordial power spectrum. II. Constraints from Planck data. {\em Physical Review D}. \textbf{89}, 1-6 (2014)

\bibitem{Meerburg2016JointEstimation} Meerburg, P., Münchmeyer, M. \& Wandelt, B. Joint resonant CMB power spectrum and bispectrum estimation. {\em Physical Review D}. \textbf{93}, 1-18 (2016)

\bibitem{Fergusson2015CombiningFeatures} Fergusson, J., Gruetjen, H., Shellard, E. \& Liguori, M. Combining power spectrum and bispectrum measurements to detect oscillatory features. {\em Physical Review D}. \textbf{91} (2015)

\bibitem{Fergusson2015PolyspectraData} Fergusson, J., Gruetjen, H., Shellard, E. \& Wallisch, B. Polyspectra searches for sharp oscillatory features in cosmic microwave sky data. {\em Physical Review D}. \textbf{91} (2015,12), http://arxiv.org/abs/1412.6152%20http://dx.doi.org/10.1103/PhysRevD.91.123506

\bibitem{Hu2015SearchingSpectra} Hu, B. \& Torrado, J. Searching for primordial localized features with CMB and LSS spectra. {\em Physical Review D}. \textbf{91}, 064039 (2015,3), https://link.aps.org/doi/10.1103/PhysRevD.91.064039

\bibitem{Bridle2003ReconstructingSpectrum} Bridle, S., Lewis, A., Weller, J. \& Efstathiou, G. Reconstructing the primordial power spectrum. {\em Monthly Notices Of The Royal Astronomical Society}. \textbf{342}, L72-L78 (2003,7), https://academic.oup.com/mnras/article-lookup/doi/10.1046/j.1365-8711.2003.06807.x

\bibitem{Bridges2009BayesianSpectrum} Bridges, M., Feroz, F., Hobson, M. \& Lasenby, A. Bayesian optimal reconstruction of the primordial power spectrum. {\em Monthly Notices Of The Royal Astronomical Society}. \textbf{400}, 1075-1084 (2009,12), https://academic.oup.com/mnras/article-lookup/doi/10.1111/j.1365-2966.2009.15525.x

\bibitem{Sealfon2005spline} Sealfon, C., Verde, L. \& Jimenez, R. Smoothing spline primordial power spectrum reconstruction. {\em Physical Review D}. \textbf{72}, 103520 (2005,11), https://link.aps.org/doi/10.1103/PhysRevD.72.103520

\bibitem{Verde2008spline} Verde, L. \& Peiris, H. On minimally parametric primordial power spectrum reconstruction and the evidence for a red tilt. {\em Journal Of Cosmology And Astroparticle Physics}. \textbf{2008}, 009 (2008,7), https://iopscience.iop.org/article/10.1088/1475-7516/2008/07/009

\bibitem{Peiris2010spline} Peiris, H. \& Verde, L. The shape of the primordial power spectrum: A last stand before Planck data. {\em Physical Review D}. \textbf{81}, 021302 (2010,1), https://link.aps.org/doi/10.1103/PhysRevD.81.021302

\bibitem{Vazquez2012ModelSpectrum} Vázquez, J., Bridges, M., Hobson, M. \& Lasenby, A. Model selection applied to reconstruction of the Primordial Power Spectrum. {\em Journal Of Cosmology And Astroparticle Physics}. \textbf{2012}, 006-006 (2012,6), https://iopscience.iop.org/article/10.1088/1475-7516/2012/06/006

\bibitem{Aslanyan2014TheSpectrum} Aslanyan, G., Price, L., Abazajian, K. \& Easther, R. The Knotted Sky I: Planck constraints on the primordial power spectrum. {\em Journal Of Cosmology And Astroparticle Physics}. \textbf{2014}, 052-052 (2014,8), https://iopscience.iop.org/article/10.1088/1475-7516/2014/08/052

\bibitem{Abazajian2014TheSpectrum} Abazajian, K., Aslanyan, G., Easther, R. \& Price, L. The Knotted Sky II: does BICEP2 require a nontrivial primordial power spectrum?. {\em Journal Of Cosmology And Astroparticle Physics}. \textbf{2014}, 053-053 (2014,8), https://iopscience.iop.org/article/10.1088/1475-7516/2014/08/053

\bibitem{Handley2019BayesianData} Handley, W., Lasenby, A., Peiris, H. \& Hobson, M. Bayesian inflationary reconstructions from Planck 2018 data. {\em Physical Review D}. \textbf{100}, 103511 (2019,11), https://link.aps.org/doi/10.1103/PhysRevD.100.103511

\bibitem{Leach2006PCA} Leach, S. Measuring the primordial power spectrum: principal component analysis of the cosmic microwave background. {\em Monthly Notices Of The Royal Astronomical Society}. \textbf{372}, 646-654 (2006,10), https://academic.oup.com/mnras/article-lookup/doi/10.1111/j.1365-2966.2006.10842.x

\bibitem{Wang1999CosmologyModels} Wang, Y., Spergel, D. \& Strauss, M. Cosmology in the Next Millennium: Combining Microwave Anisotropy Probe and Sloan Digital Sky Survey Data to Constrain Inflationary Models. {\em The Astrophysical Journal}. \textbf{510}, 20-31 (1999,1), https://iopscience.iop.org/article/10.1086/306558

\bibitem{Hannestad2001ReconstructingData} Hannestad, S. Reconstructing the inflationary power spectrum from cosmic microwave background radiation data. {\em Physical Review D}. \textbf{63}, 043009 (2001,1), https://link.aps.org/doi/10.1103/PhysRevD.63.043009

\bibitem{Hannestad2004ReconstructingAlgorithm} Hannestad, S. Reconstructing the primordial power spectrum—a new algorithm. {\em Journal Of Cosmology And Astroparticle Physics}. \textbf{2004}, 002-002 (2004,4), https://iopscience.iop.org/article/10.1088/1475-7516/2004/04/002

\bibitem{Elgaroy2002} Elgaroy, O., Gramann, M. \& Lahav, O. Features in the primordial power spectrum: Constraints from the cosmic microwave background and the limitation of the 2dF and SDSS redshift surveys to detect them. {\em Monthly Notices Of The Royal Astronomical Society}. \textbf{333}, 93-99 (2002,6), https://academic.oup.com/mnras/article/333/1/93/1188065

\bibitem{Matsumiya2002CosmicAnisotropy} Matsumiya, M., Sasaki, M. \& Yokoyama, J. Cosmic inversion: Reconstructing the primordial spectrum from CMB anisotropy. {\em Physical Review D}. \textbf{65}, 083007 (2002,4), https://link.aps.org/doi/10.1103/PhysRevD.65.083007

\bibitem{Matsumiya2003CosmicData} Matsumiya, M., Sasaki, M. \& Yokoyama, J. Cosmic inversion: II. An iterative method for reproducing the primordial spectrum from the CMB data. {\em Journal Of Cosmology And Astroparticle Physics}. \textbf{2003}, 003-003 (2003,2), https://iopscience.iop.org/article/10.1088/1475-7516/2003/02/003

\bibitem{Kogo2004ReconstructingMethod} Kogo, N., Matsumiya, M., Sasaki, M. \& Yokoyama, J. Reconstructing the Primordial Spectrum from WMAP Data by the Cosmic Inversion Method. {\em The Astrophysical Journal}. \textbf{607}, 32-39 (2004,5), https://iopscience.iop.org/article/10.1086/383339

\bibitem{Tegmark2002SeparatingBox} Tegmark, M. \& Zaldarriaga, M. Separating the early universe from the late universe: Cosmological parameter estimation beyond the black box. {\em Physical Review D}. \textbf{66}, 103508 (2002,11), https://link.aps.org/doi/10.1103/PhysRevD.66.103508

\bibitem{Tocchini-Valentini2005ASpectrum} Tocchini-Valentini, D., Douspis, M. \& Silk, J. A new search for features in the primordial power spectrum. {\em Monthly Notices Of The Royal Astronomical Society}. \textbf{359}, 31-35 (2005,5), https://academic.oup.com/mnras/article-lookup/doi/10.1111/j.1365-2966.2005.08771.x

\bibitem{Tocchini-Valentini2006Non-parametricData} Tocchini-Valentini, D., Hoffman, Y. \& Silk, J. Non-parametric reconstruction of the primordial power spectrum at horizon scales from WMAP data. {\em Monthly Notices Of The Royal Astronomical Society}. \textbf{367}, 1095-1102 (2006,4), https://academic.oup.com/mnras/article-lookup/doi/10.1111/j.1365-2966.2006.10031.x

\bibitem{Nagata2009Band-powerMethod} Nagata, R. \& Yokoyama, J. Band-power reconstruction of the primordial fluctuation spectrum by the maximum likelihood reconstruction method. {\em Physical Review D}. \textbf{79}, 043010 (2009,2), https://link.aps.org/doi/10.1103/PhysRevD.79.043010

\bibitem{Hunt2014ReconstructionSets} Hunt, P. \& Sarkar, S. Reconstruction of the primordial power spectrum of curvature perturbations using multiple data sets. {\em Journal Of Cosmology And Astroparticle Physics}. \textbf{2014}, 025-025 (2014,1), https://iopscience.iop.org/article/10.1088/1475-7516/2014/01/025

\bibitem{Hunt2015SearchDatasets} Hunt, P. \& Sarkar, S. Search for features in the spectrum of primordial perturbations using Planck and other datasets. {\em Journal Of Cosmology And Astroparticle Physics}. \textbf{2015}, 052-052 (2015,12), https://iopscience.iop.org/article/10.1088/1475-7516/2015/12/052

\bibitem{Gauthier2012ReconstructingCMB} Gauthier, C. \& Bucher, M. Reconstructing the primordial power spectrum from the CMB. {\em Journal Of Cosmology And Astroparticle Physics}. \textbf{2012}, 050-050 (2012,10), https://iopscience.iop.org/article/10.1088/1475-7516/2012/10/050

\bibitem{Mukherjee2003DirectSpectrum} Mukherjee, P. \& Wang, Y. Direct Wavelet Expansion of the Primordial Power Spectrum. {\em The Astrophysical Journal}. \textbf{598}, 779-784 (2003,12), https://iopscience.iop.org/article/10.1086/379107

\bibitem{Mukherjee2005PrimordialReconstruction} Mukherjee, P. \& Wang, Y. Primordial power spectrum reconstruction. {\em Journal Of Cosmology And Astroparticle Physics}. \textbf{2005}, 007-007 (2005,12), https://iopscience.iop.org/article/10.1088/1475-7516/2005/12/007

\bibitem{Nicholson2010ReconstructionInversion} Nicholson, G., Contaldi, C. \& Paykari, P. Reconstruction of the primordial power spectrum by direct inversion. {\em Journal Of Cosmology And Astroparticle Physics}. \textbf{2010}, 016-016 (2010,1), https://iopscience.iop.org/article/10.1088/1475-7516/2010/01/016

\bibitem{Shafieloo2004MRL} Shafieloo, A. \& Souradeep, T. Primordial power spectrum from WMAP. {\em Physical Review D}. \textbf{70}, 043523 (2004,8), https://link.aps.org/doi/10.1103/PhysRevD.70.043523

\bibitem{Shafieloo2007FeaturesAnalysis} Shafieloo, A., Souradeep, T., Manimaran, P., Panigrahi, P. \& Rangarajan, R. Features in the primordial spectrum from WMAP: A wavelet analysis. {\em Physical Review D}. \textbf{75}, 123502 (2007,6), https://link.aps.org/doi/10.1103/PhysRevD.75.123502

\bibitem{Shafieloo2008EstimationData} Shafieloo, A. \& Souradeep, T. Estimation of primordial spectrum with post-WMAP 3-year data. {\em Physical Review D}. \textbf{78}, 023511 (2008,7), https://link.aps.org/doi/10.1103/PhysRevD.78.023511

\bibitem{Nicholson2009ReconstructionExperiments} Nicholson, G. \& Contaldi, C. Reconstruction of the primordial power spectrum using temperature and polarisation data from multiple experiments. {\em Journal Of Cosmology And Astroparticle Physics}. \textbf{2009}, 011-011 (2009,7), https://iopscience.iop.org/article/10.1088/1475-7516/2009/07/011

\bibitem{Gibelyou2010DetectabilityDistribution} Gibelyou, C., Huterer, D. \& Fang, W. Detectability of large-scale power suppression in the galaxy distribution. {\em Physical Review D}. \textbf{82}, 123009 (2010,12), https://link.aps.org/doi/10.1103/PhysRevD.82.123009

\bibitem{Hamann2010FeaturesAnalysis} Hamann, J., Shafieloo, A. \& Souradeep, T. Features in the primordial power spectrum? A frequentist analysis. {\em Journal Of Cosmology And Astroparticle Physics}. \textbf{2010}, 010-010 (2010,4), https://iopscience.iop.org/article/10.1088/1475-7516/2010/04/010

\bibitem{Hazra2013wmapMRL} Hazra, D., Shafieloo, A. \& Souradeep, T. Primordial power spectrum: a complete analysis with the WMAP nine-year data. {\em Journal Of Cosmology And Astroparticle Physics}. \textbf{2013}, 031-031 (2013,7), https://iopscience.iop.org/article/10.1088/1475-7516/2013/07/031

\bibitem{Hazra2013CosmologicalSpectrum} Hazra, D., Shafieloo, A. \& Souradeep, T. Cosmological parameter estimation with free-form primordial power spectrum. {\em Physical Review D}. \textbf{87}, 123528 (2013,6), https://link.aps.org/doi/10.1103/PhysRevD.87.123528

\bibitem{Hazra2014PlanckMRL} Hazra, D., Shafieloo, A. \& Souradeep, T. Primordial power spectrum from Planck. {\em Journal Of Cosmology And Astroparticle Physics}. \textbf{2014}, 011-011 (2014,11), https://iopscience.iop.org/article/10.1088/1475-7516/2014/11/011

\bibitem{Chandra2021PrimordialSpectrum} Chandra, R. \& Souradeep, T. Primordial Power Spectrum reconstruction from CMB Weak Lensing Power Spectrum. {\em Journal Of Cosmology And Astroparticle Physics}. \textbf{2021}, 081 (2021)

\bibitem{Richardson1972Bayesian-BasedRestoration} Richardson, W. Bayesian-Based Iterative Method of Image Restoration. {\em Journal Of The Optical Society Of America}. \textbf{62}, 55 (1972,1)

\bibitem{Lucy1974AnDistributions} Lucy, L. An iterative technique for the rectification of observed distributions. {\em The Astronomical Journal}. \textbf{79} (1974,6)

\bibitem{Baugh1993TheFunction} Baugh, C. \& Efstathiou, G. The Three-Dimensional Power Spectrum Measured from the APM Galaxy Survey - Part One - Use of the Angular Correlation Function. {\em Monthly Notices Of The Royal Astronomical Society}. \textbf{265}, 145 (1993,11)

\bibitem{Baugh1994TheSpectrum} Baugh, C. \& Efstathiou, G. The three-dimensional power spectrum measured from the APM Galaxy Survey-2. Use of the two-dimensional power spectrum. {\em Monthly Notices Of The Royal Astronomical Society}. \textbf{267}, 323-332 (1994)

\bibitem{Katsaggelos1991AAlgorithm} Katsaggelos, A., Biemond, J., Schafer, R. \& Mersereau, R. A regularized iterative image restoration algorithm. {\em IEEE Transactions On Signal Processing}. \textbf{39}, 914-929 (1991,4)

\bibitem{Conchello1996FastMicroscopy} Conchello, J. \& McNally, J. Fast regularization technique for expectation maximization algorithm for optical sectioning microscopy. {\em Three-Dimensional Microscopy: Image Acquisition And Processing III}. pp. 199-208 (1996,4)

\bibitem{Dey2004ARegularization} Dey, N., Blanc-Feraud, L., Zimmer, C., Kam, Z., Olivo-Marin, J. \& Zerubia, J. A deconvolution method for confocal microscopy with total variation regularization. {\em 2004 2nd IEEE International Symposium On Biomedical Imaging: Macro To Nano (IEEE Cat No. 04EX821)}. pp. 1223-1226 (2004)

\bibitem{Lewis2000camb} Lewis, A., Challinor, A. \& Lasenby, A. Efficient Computation of Cosmic Microwave Background Anisotropies in Closed Friedmann‐Robertson‐Walker Models. {\em The Astrophysical Journal}. \textbf{538}, 473-476 (2000), http://stacks.iop.org/0004-637X/538/i=2/a=473

\bibitem{Blas2011class} Blas, D., Lesgourgues, J. \& Tram, T. The Cosmic Linear Anisotropy Solving System (CLASS). Part II: Approximation schemes. {\em Journal Of Cosmology And Astroparticle Physics}. \textbf{2011}, 034-034 (2011,7), https://iopscience.iop.org/article/10.1088/1475-7516/2011/07/034

\bibitem{PlanckCollaboration2018power} Planck Collaboration Planck 2018 results. V. CMB power spectra and likelihoods. {\em Astronomy \& Astrophysics}. \textbf{641} pp. A5 (2020)

\bibitem{Efstathiou2021AMaps} Efstathiou, G. \& Gratton, S. A Detailed Description of the CAMSPEC Likelihood Pipeline and a Reanalysis of the Planck High Frequency Maps. {\em The Open Journal Of Astrophysics}. \textbf{4} (2021,8), https://astro.theoj.org/article/27518-a-detailed-description-of-the-camspec-likelihood-pipeline-and-a-reanalysis-of-the-planck-high-frequency-maps

\bibitem{Powell2009TheDerivatives.} Powell, M. The BOBYQA algorithm for bound constrained optimization without derivatives.. {\em Cambridge NA Report NA2009/06, University Of Cambridge, Cambridge }. \textbf{26} (2009)

\bibitem{PlanckCollaboration2016PlanckParameters} Planck Collaboration Planck 2015 results. XI. CMB power spectra, likelihoods, and robustness of cosmological parameters. {\em Astronomy \& Astrophysics}. \textbf{594} pp. A11 (2016,10), http://www.aanda.org/10.1051/0004-6361/201526926

\bibitem{Hazra2014InflationSpectra} Hazra, D., Shafieloo, A., Smoot, G. \& Starobinsky, A. Inflation with Whip-Shaped Suppressed Scalar Power Spectra. {\em Physical Review Letters}. \textbf{113}, 071301 (2014,8), https://link.aps.org/doi/10.1103/PhysRevLett.113.071301

\bibitem{Jain2009PunctuatedMultipoles} Jain, R., Chingangbam, P., Gong, J., Sriramkumar, L. \& Souradeep, T. Punctuated inflation and the low CMB multipoles. {\em Journal Of Cosmology And Astroparticle Physics}. \textbf{2009}, 009-009 (2009,1), https://iopscience.iop.org/article/10.1088/1475-7516/2009/01/009

\bibitem{Contaldi2003SuppressingAnisotropies} Contaldi, C., Peloso, M., Kofman, L. \& Linde, A. Suppressing the lower multipoles in the CMB anisotropies. {\em Journal Of Cosmology And Astroparticle Physics}. \textbf{2003}, 002-002 (2003,7), https://iopscience.iop.org/article/10.1088/1475-7516/2003/07/002

\bibitem{Hergt2019CaseInflation} Hergt, L., Handley, W., Hobson, M. \& Lasenby, A. Case for kinetically dominated initial conditions for inflation. {\em Physical Review D}. \textbf{100}, 023502 (2019,7), https://link.aps.org/doi/10.1103/PhysRevD.100.023502

\bibitem{Ragavendra2022SuppressionBispectra} Ragavendra, H., Chowdhury, D. \& Sriramkumar, L. Suppression of scalar power on large scales and associated bispectra. {\em Physical Review D}. \textbf{106}, 043535 (2022,8), https://link.aps.org/doi/10.1103/PhysRevD.106.043535

\bibitem{Hazra2010PrimordialPotential} Hazra, D., Aich, M., Jain, R., Sriramkumar, L. \& Souradeep, T. Primordial features due to a step in the inflaton potential. {\em Journal Of Cosmology And Astroparticle Physics}. \textbf{2010}, 008 (2010)

\bibitem{Chluba2015} Chluba, J., Hamann, J. \& Patil, S. Features and new physical scales in primordial observables: Theory and observation. {\em International Journal Of Modern Physics D}. \textbf{24}, 1530023 (2015,9), https://www.worldscientific.com/doi/abs/10.1142/S0218271815300232

\bibitem{Miranda2016NonlinearSpectra} Miranda, V., Hu, W., He, C. \& Motohashi, H. Nonlinear excitations in inflationary power spectra. {\em Physical Review D}. \textbf{93}, 023504 (2016,1), https://link.aps.org/doi/10.1103/PhysRevD.93.023504

\bibitem{Hazra2014WigglyInflation} Hazra, D., Shafieloo, A., Smoot, G. \& Starobinsky, A. Wiggly whipped inflation. {\em Journal Of Cosmology And Astroparticle Physics}. \textbf{2014}, 048-048 (2014,8), https://iopscience.iop.org/article/10.1088/1475-7516/2014/08/048

\bibitem{Hazra2016PrimordialPolarization} Hazra, D., Shafieloo, A., Smoot, G. \& Starobinsky, A. Primordial features and Planck polarization. {\em Journal Of Cosmology And Astroparticle Physics}. \textbf{1609}, 9 (2016)

\bibitem{Hazra2018ProbingObservations} Hazra, D., Paoletti, D., Ballardini, M., Finelli, F., Shafieloo, A., Smoot, G. \& Starobinsky, A. Probing features in inflaton potential and reionization history with future CMB space observations. {\em Journal Of Cosmology And Astroparticle Physics}. \textbf{2018}, 017 (2018)

\bibitem{Hazra2021InflationBeyond} Hazra, D., Paoletti, D., Debono, I., Shafieloo, A., Smoot, G. \& Starobinsky, A. Inflation story: slow-roll and beyond. {\em Journal Of Cosmology And Astroparticle Physics}. \textbf{12}, 38 (2021)

\bibitem{Chen2011FingerprintsPerturbations} Chen, X. Fingerprints of primordial universe paradigms as features in density perturbations. {\em Physics Letters B}. \textbf{706}, 111-115 (2011,12), https://linkinghub.elsevier.com/retrieve/pii/S0370269311013621

\bibitem{Chen2012PrimordialInflation} Chen, X. Primordial features as evidence for inflation. {\em Journal Of Cosmology And Astroparticle Physics}. \textbf{2012}, 038-038 (2012,1), https://iopscience.iop.org/article/10.1088/1475-7516/2012/01/038

\bibitem{Chen2015ModelsClock} Chen, X., Namjoo, M. \& Wang, Y. Models of the Primordial Standard Clock. {\em Journal Of Cosmology And Astroparticle Physics}. \textbf{2015}, 027-027 (2015,2), https://iopscience.iop.org/article/10.1088/1475-7516/2015/02/027

\bibitem{Braglia2021ComparingData} Braglia, M., Chen, X. \& Hazra, D. Comparing multi-field primordial feature models with the Planck data. {\em Journal Of Cosmology And Astroparticle Physics}. \textbf{2021}, 005 (2021,6), https://iopscience.iop.org/article/10.1088/1475-7516/2021/06/005

\bibitem{Braglia2022PrimordialAnomalies} Braglia, M., Chen, X. \& Hazra, D. Primordial standard clock models and CMB residual anomalies. {\em Physical Review D}. \textbf{105}, 103523 (2022,5), https://link.aps.org/doi/10.1103/PhysRevD.105.103523

\bibitem{Braglia2022UncoveringSpectra} Braglia, M., Chen, X. \& Hazra, D. Uncovering the history of cosmic inflation from anomalies in cosmic microwave background spectra. {\em The European Physical Journal C}. \textbf{82}, 498 (2022,5), https://link.springer.com/10.1140/epjc/s10052-022-10461-3

\bibitem{Antony2022DiscordancesDynamics} Antony, A., Finelli, F., Hazra, D. \& Shafieloo, A. Discordances in cosmology and the violation of slow-roll inflationary dynamics.  (2022,2), http://arxiv.org/abs/2202.14028

\bibitem{PlanckCollaboration2013likelihood} Planck Collaboration Planck 2013 results. XV. CMB power spectra and likelihood. {\em Astronomy \& Astrophysics}. \textbf{572} pp. A15 (2014)





\end{thebibliography}
\end{document}